\documentstyle[epsfig]{aipproc}

\begin{document}
\title{Phaseseparation in overdoped 
Y$_{1-0.8}$Ca$_{0-0.2}$Ba$_2$Cu$_3$O$_{6.96-6.98}$}
\author{\mbox{J. R\"ohler$^*$\thanks{Corresponding author:
abb12@rs1.rrz.uni-koeln.de}, C. Friedrich$^*$, T. Granzow$^*$, 
E. Kaldis$^{\dagger}$, and G. B\"ottger$^{\dagger}$}}
\address{$^*$Universit\"at zu K\"oln, D-50937 K\"oln, Germany\\
$^{\dagger}$ETH Z\"urich, CH-8093 Z\"urich, Switzerland}
\maketitle
\begin{abstract} 
The\nolinebreak[4] dimpling\nolinebreak[4] in\nolinebreak[4] 
the\nolinebreak[4]\nolinebreak[4] CuO$_2$\nolinebreak[4] planes
\nolinebreak[4] of\nolinebreak[4] overdoped\nolinebreak[4] 
Y$_{1-y}$Ca$_y$Ba$_2$Cu$_3$\linebreak[4] O$_{6.96-6.98}$, 
($y=0.02-0.2$) has been measured by x-ray absorption-fine-structure
spectroscopy (Y-$K$ EXAFS). A step-like decrease around 12\% Ca 
indicates a percolation threshold for distorted sites of 5 cells, 
and thus phase segregation. We conclude the 
charge carriers added by substitution of Y$^{3+}$ by Ca$^{2+}$ to be
trapped at the Ca sites and their {\it nn} environment.
\end{abstract}

\section*{Introduction}
The unusual metallic properties of the high $T_c$\/ cuprate 
superconductors are difficult to reconcile with a homogenous 
electronic state. Since inhomogeneities in the electronic structure 
may lift the translational invariance of the underlying lattice, 
it is suggesting to measure both, the atomic structure using 
short-range (or local) structural probes, and the average 
crystallographic structure using diffraction techniques 
\cite{RoeKal}. Anomalous atomic displacements may be then 
extracted from careful comparisons between the local and the 
crystallographic structure. 

The anomalous electronic structure 
of the cuprate supercoductors is frequently discussed in terms 
of dynamic inhomogenieties, for instance a mixture of 
microscopically segregated phases. 
The notorious nonstoichiometry of all known 
superconducting cuprates, even their 
optimum doped phases, causes many static inhomogeneities thus 
adding a constraint to the analysis of 
anomalous atomic displacements.

Advantageously the metallic CuO$_2$\/-planes of the
cuprate superconductors are the 
structurally most perfect blocks. Thus significant structural 
anomalies in the planes can be safely related to nontrivial 
electronic inhomogeneities. We have recently shown 
that upon oxygen doping of YBa$_2$Cu$_3$O$_{x}$ 
the locally measured spacing between the Cu2 and O2,3 
layers (``dimpling``) in the 
CuO$_2$\/-planes keeps track of the planar hole 
concentration \cite{RoeKal}. The data from 
Y $K$\/-edge EXAFS comprise the atomic 
structure of the 8 next neighboured 
CuO$_2$\/-plaquettes, and thus the effective charge in only 32 
Cu2--O23 bonds. Oxygen doping for $x\rightarrow x_{opt}$\/=6.92 
has been found to increase the dimpling, in other words: the increasing
number of oxygen holes bends the Cu2--O23 bonds out-of-plane
towards the Ba-layer. At the onset of the overdoped regime, 
$x\simeq 6.95$, the dimpling and thus also the number of holes 
exhibits a sharp maximum \cite{KalLoe}. A concomitant displacive 
transformation of the crystallographic structure however 
seems to block a further increase of the dimpling. Although 
further doping from $x\simeq 6.95\rightarrow 7$ increases 
the nomimal hole concentration, the dimpling starts 
to decrease. 

In this contribution we report
on Y $K$\/-EXAFS measurements of the dimpling 
in a series of overdoped compounds, YBa$_2$Cu$_3$O$_{6.96-6.98}$, 
additionally overdoped by substitution of Y$^{3+}$ with 
2--20\% Ca$^{2+}$. 

\begin{figure}[h] 
\centerline{\epsfig{file=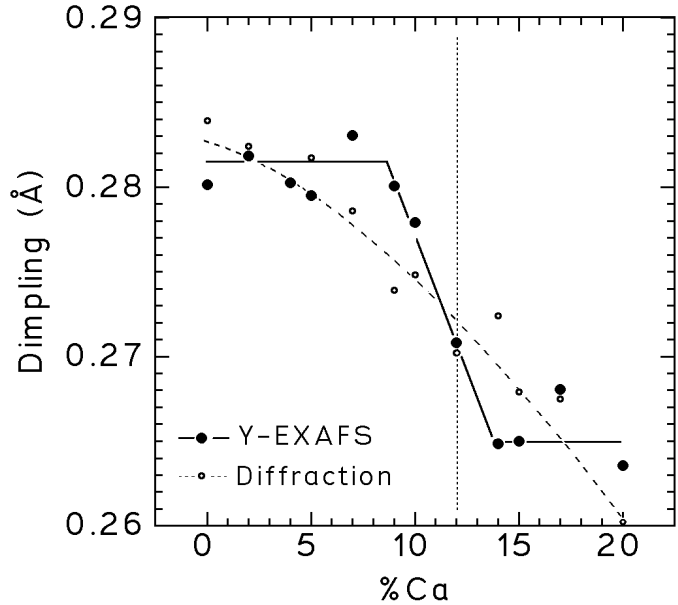,height=8.1175cm,
width=8.925cm}}
\vspace{10pt}
\caption{Dimpling of the CuO$_2$\/-planes in
Y$_{1-y}$Ca$_y$Ba$_2$Cu$_3$O$_{6.96-6.98}$ as a function of Ca
concentration. Large full circles (thick drawn out line): data from
Y-EXAFS at 25 K. Thin open circles (thin dashed line): data from
neutron diffraction at 5 K [2]. For convenience 
the latter are offset by +0.01 {\AA}.} 
\label{fig1}
\end{figure}

\section*{Experimental Details}
The polycrystalline samples were from the same batches studied 
previously by diffraction with x-rays and neutrons, and by 
magnetometry \cite{BoeFau}. Up to 20\% Ca could be homogenously 
solved in YBa$_2$Cu$_3$O$_{x}$ while the oxygen content was kept 
at the highest numbers possible: $x=6.96-6.98$, $i.e.$\/ 
surprisingly always $<7.00$. Ca EXAFS of the same samples confirmed  
that calcium has replaced yttrium ($>97$\%), and not 
barium. The EXAFS spectra ($T=20-60 K$) 
were recorded at the European Synchrotron Radation Facility (ESRF) 
using the double crystal spectrometer at BM29. Details of the 
spectroscopic technique and of the data analysis are given in 
\cite{RoeKal}.

\section*{Results}
Fig.\ref{fig1} exhibits the dimpling of the CuO$_2$\/-planes 
in Y$_{1-y}$Ca$_y$Ba$_2$Cu$_3$O$_{6.96-6.98}$ ($T=25$ K) 
for $y=0.02-0.2$ as extracted from the Y EXAFS 
(large full circles, drawn out line). Comparison is made with the 
results from  neutron diffraction (small circles, dashed line) by 
B\"ottger {\it et al.} \cite{BoeFau}. From the 
Y EXAFS the dimpling is found   
independent on the Ca concentration up to 9\% (0.281(2) {\AA}), 
then undergoes a step-like decrease by about 0.015 $\rm\AA$, and 
flatens around
0.265(4) {\AA} for 14--20 \% Ca. The position of the 
step may be located around 12\% Ca
(dotted vertical line). The discontinous behaviour of the dimpling 
from the Y EXAFS is at variance with the continous behaviour 
obtained from the refinement of the average 
crystallographic structure \cite{BoeFau}. For better comparison the 
diffraction data are offset by +0.01 {\AA} showing that both methods
yield the same overall variations, and
that the discontinuity from the EXAFS work is clearly outside the
scatter of the data points from the diffraction work.

\begin{figure}[h] 
\centerline{\epsfig{file=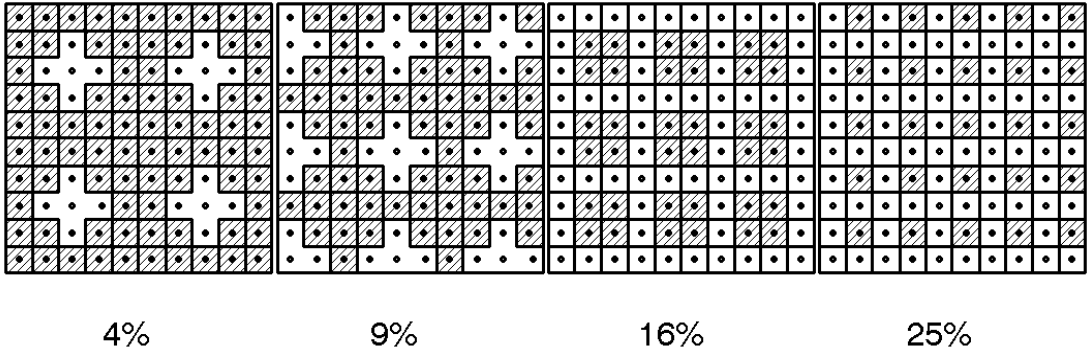,height=4.7813cm,width=14.6042cm}}
\vspace{10pt}
\caption{Square lattices of Cu2 with Y/Ca (dots) in their centers. 
Increasing concentrations (from left to right) of ordered Ca 
substitute Y. The divalent Ca impurities are 
assumed  to be screened only by the {\it nn} Y$^{3+}$. The thus 
locally distorted areas appear as white crosses starting to  
percolate somewhere between 9\% and 16\%Ca. 
The remaining undistorted Y-cells are hatched.} 
\label{fig2}
\end{figure}

\section*{Discussion}
The step-like variation of the dimpling with increasing 
Ca concentration points to a percolative transition induced by 
Ca doping. Fig. \ref{fig2} exhibits a plausible scheme 
demonstrating the occurence of percolative paths for 
concentrations around 16\% Ca. Here it is 
assumed that the holes doped by Ca$^{2+}$ 
are predominantly screened by the {\it nn} Y cells 
thus creating a cross-like cluster of 5 distorted cells. 
It is suggesting that these holes are trapped and do not 
contribute to the density of the mobile carriers.  

Considering the Ca impurities  
as percolating sites in a random process, 
the exact theory for a square 2-D lattice 
\cite{StaAha} predicts the critical 
percolation to occur for 59.2746\%. 
Then straightforwardly the critical percolation for 
the cross-like clusters, each centered at a Ca impurity,  
is expected for $59.2746 \%/5 \simeq 12$\%  Ca. We conclude that 
the observed step-like decrease of the dimpling around 
12\% Ca is connected to this percolation threshold. 

The other way around: the observation of a percolation 
threshold at 12\% Ca indicates the percolating 
sites to be 5 cells large. The solid solution thus segregates 
into two phases: 
{\it i.} the matrix of undistorted Y cells, and {\it ii.} distorted 
clusters at the Ca sites.

\section*{Concluding Remarks}
We conclude that doping of YBa$_2$Cu$_3$O$_x$ with heterovalent 
cations substituting Y is electronically 
nonequivalent with oxygen doping in the chain layer.
Thus generalized phase diagrams treating Ca$^{+2}$ and O$^{-2}$  
dopants in Y$_{1-y}$Ca$_y$Ba$_2$Cu$_3$O$_x$ 
on an equal footing are questionable. 
In particular the superconducting 
transition temperature of dually doped 
Y$_{1-y}$Ca$_y$Ba$_2$Cu$_3$O$_x$ is not 
expected to match a parabolic behaviour $T_c$  {\it vs.} 
hole concentration.

\section*{Acknowledgments}
Beamtime at the ESRF was under the proposals HS376 and HS 533.
We thank S. Thienhaus for help during the data acquisition.


\begin{references}

\bibitem{RoeKal} R\"ohler, J., Loeffen, P. W., M\"ullender, S., 
Conder, K., and Kaldis, E.,  in {\it Workshop on High-T$_c$
Superconductivity 1996: Ten Years after the Discovery}, 
Kaldis, E., Liarokapis, E., and M\"uller, K. A., eds., NATO ASI 
Series E343, Dordrecht, Kluwer, 1997, pp. 469--502;
cond-mat/9707208.\\

\bibitem{BoeFau} B\"ottger, G., Mangelshots, I., Kaldis, E., Fischer,
P., Kr\"uger, Ch., and Fauth, F., {\it J. Phys: Cond. Matter}
{\bf 8}, 8889 (1996).\\

\bibitem{KalLoe}Kaldis, E., R\"ohler, J., Liarokapis, E.,
Poulakis, N., Conder, K., and Loeffen, P. W., {\it Phys. Rev.
Lett.} {\bf 79}, 4894--4897 (1997); 
cond-mat/9707196.\\

\bibitem{StaAha}Stauffer, D., and Aharony A., 
{\it Perkolationstheorie:  eine Einf\"uhrung}, Weinheim, VCH, 
1995, ch. 2, p. 18.

\end{references}
\end{document}